\documentclass[aps,preprint,showpacs,showkeys]{revtex4}   
\usepackage{epsfig} 
\usepackage{ amsmath,amssymb,amsthm}    
\usepackage{graphicx}
\usepackage{psfrag}
\usepackage{bm}
\usepackage[dvips]{color}   
\newcounter{egyenlet}[equation]   
   
\begin{document}   
   
\renewcommand{\thesection}{\arabic{section}}   
\renewcommand{\theequation}{\arabic{equation}\alph{egyenlet}}   
\def\dl{\displaystyle}    
\def\arrowlimit#1{\mathrel{\mathop{\longrightarrow}\limits_{#1}}}   
\def\grad{{\rm grad}}   
\def\div{{\rm div}}   
\def\rot{{\rm rot}}   
\def\scs{\scriptsize}   
\def\l{\langle}   
\def\r{\rangle}

\title{\Large Generalized statistical complexity and \\   
Fisher-R\'enyi entropy product in the $H$-atom}   
   
\author{\bf E. Romera$^1$, R. L\'opez-Ruiz$^2$, J. Sa\~nudo$^3$,  and \'A. Nagy$^4$}  
\affiliation{  
\small $^1$Departamento de F\'{\i}sica At\'omica, Molecular y Nuclear and Instituto   
Carlos I de F{\'\i}sica Te\'orica y Computacional,   
Universidad de Granada, E-18071 Granada, Spain \\   
$^2$DIIS and BIFI, Facultad de Ciencias, Universidad de Zaragoza, E-50009 Zaragoza, Spain\\  
$^3$Departamento de F\'{\i}sica, Facultad de Ciencias, Universidad de Extremadura,   
E-06071 Badajoz, Spain\\  
$^4$Department of Theoretical Physics, University of Debrecen, H-4010 Debrecen, Hungary}   
\date{\today}

\begin{abstract}   
By using the R\'enyi entropy,  
and following the same scheme that in the Fisher-R\'enyi entropy product case, 
a generalized statistical complexity is defined.  
Several properties of it, including  inequalities and lower and upper bounds are derived. 
The hydrogen atom is used as a test system where to quantify these two different statistical  
magnitudes, the Fisher-R\'enyi entropy product and the generalized statistical complexity.  
For each level of energy, both indicators take their minimum values on the orbitals that correspond  
to the highest orbital angular momentum. Hence, in the same way as happens with the Fisher-Shannon  
and the statistical complexity, these generalized R\'enyi-like statistical magnitudes  
break the energy degeneration in the H-atom.   
\end{abstract}  
  
\maketitle  
\section{Introduction}  
  
Nowadays the study of statistical magnitudes in quantum systems  
has a role of growing importance. So, information entropies   
and statistical complexities have been calculated on different  
atomic systems \cite{gadre1985,panos2005}. In particular, the $H$-atom is a natural test system   
where to quantify all this kind of magnitudes \cite{dehesa1994,coffey2003,romera2005,szabo2008,sanudo2008}.  
This system displays a remarkable property when Fisher-Shannon information \cite{vignat2003,romera2004,angulo2008}   
and the so called LMC complexity \cite{lopez1995,catalan2002} are computed on it.   
Namely, for each energy level, the minimum values of those statistical measures are taken on the   
wave functions with the highest orbital angular momentum, 
those orbitals that correspond to the Bohr-like orbits in the pre-quantum image \cite{sanudo2008}.  
  
Shannon information \cite{shannon1948} plays an important role in both entropic products,  
the Fisher-Shannon information and the LMC statistical complexity.  
Different generalizations of the Shannon information dependent on  
a parameter $\alpha$ can be found in the literature \cite{renyi1961,tsallis1988}.  
Implementing these $\alpha$-entropies in both entropic products,  
different families of statistical indicators can be generated \cite{rosso2006}.  
In particular, if R\'enyi entropies are taken in exponential form for this purpose,   
the Fisher-R\'enyi entropy product \cite{romera2008} and   
a new generalized LMC statistical complexity can be defined.   
  
In this work, an information theoretical analysis of the same  
quantum system, the $H$-atom, is presented in terms of the Fisher-R\'enyi entropy product and the  
generalized LMC statistical measure. These $\alpha$-dependent magnitudes show a similar behavior to   
that found in the limit $\alpha\rightarrow 1$, that correspond to the already studied case of the  
Fisher-Shannon information and LMC complexity \cite{sanudo2008}. That is, the degeneration of the   
energy is also broken by these statistical magnitudes.     
  
The paper is organized as follows. In section 2, the Fisher-R\'enyi entropy product is recalled   
and the generalized LMC complexity measure is introduced.  
Some of their properties are also presented. The calculation of these magnitudes  
in the $H$-atom is presented in section 3. Our conclusions are contained in the last section.

\section{Complexity Measures}  
  
Consider a $D$-dimensional distribution function $f({\bf r})$, with   
$f({\bf r})$ nonnegative and   
{$\int f({\bf r}) d{\bf r}=1$  and   
define   R\'enyi entropy power of index $\alpha$ blue as in 
  \cite{romera2008} by   
\begin{equation}  
N_{f}^{(\alpha)} =  
\frac{\eta_{\alpha}}{2\pi}\exp{\left(\frac{2}{D}R_f^{(\alpha)}\right)},  
\quad\quad \text{with}\quad \quad  \alpha>1/2,  
\label{r1b}  
\end{equation}  
being $\eta_{\alpha}=\left(\frac{\alpha}{2\alpha-1}\right)^{\frac{2\alpha-1}{\alpha-1}}$  
a decreasing function from $1$ to $0$ when $\alpha$ runs in $(0.5,\infty)$ and  
$\eta_{\alpha=1}=e^{-1}$, and the R\'enyi entropy of order $\alpha$ given by  
\begin{equation}  
R_f^{(\alpha)} = \frac{1}{1-\alpha}\ln\int [f({\bf r})]^{\alpha} d{\bf r},  
\end{equation}  
where $\bf r$ stands for $r_1, ..., r_D$.  
  
R\'enyi entropy power is an extension of Shannon entropy power \cite{dembo1991} and  
verifies that when $\alpha\rightarrow 1$ then $N_f^{(\alpha)}\rightarrow  
N_f=\frac{1}{2\pi e}e^{\frac{2}{D}S_f}$ with  
\begin{equation}  
S_f = -\int f({\bf r})\ln f({\bf r}) d{\bf r} .  
\end{equation}  
A scaling property is verified by R\'enyi entropy power \cite{romera2008}, which transforms as  
\begin{equation}  
N_{|\Psi_\lambda|^2}^{(\alpha)}=\lambda^{-2} N_{|\Psi|^2}^{(\alpha)},  
\label{scalingrenyi}  
\end{equation}  
under scaling of the function $\Psi_{\lambda}(r_1,...,r_D)=\lambda^{D/2}  
\Psi(\lambda r_1,...,\lambda r_D)$.  
R\'enyi entropy power also has the property \cite{romera2008}  
\begin{equation}  
N_{f}^{(\alpha)} > N_f^{(\alpha')} \quad \mbox{for}\quad \alpha<\alpha' .  
\label{r1a}  
\end{equation}  
Fisher information \cite{fisher1925} of the probability density function $f$ is given by  
\begin{equation}  
I_f = \int \frac{|\nabla f({\bf r})|^2}{f({\bf r})}d{\bf r}.  
\end{equation}   
  
\subsection{Fisher-R\'enyi Entropy Product $P_f^{(\alpha)}$}  
  
The Fisher-R\'enyi entropy product is defined by   
\begin{equation}  
P_f^{(\alpha)}=\frac1D N_f^{(\alpha)} I_f, \quad  \mbox{with} \quad \quad \alpha\in(1/2,1].  
\end{equation}  
It displays the following important properties \cite{romera2008}:\newline  
(i) $P_{f}^{(\alpha)}$ is invariant under  scaling transformation  
$f_{\lambda}=\lambda^{D}f(\lambda {\bf r})$, i. e.  
$P_{f_{\lambda}}^{(\alpha)}=P_f^{(\alpha)}$, \newline  
(ii) it verifies the inequality $P_f^{(\alpha)}\geq 1$, and \newline   
(iii) $P_{f}^{(\alpha)}$ is a nonincreasing function of $\alpha$ for any probability density.  
  
\subsection{Generalized Complexity Measure $C_f^{(\alpha)}$}  
  
The measure of complexity $C_{f}$ introduced in \cite{lopez1995,catalan2002}, the  
so-called LMC complexity, is defined by   
\begin{equation}  
C_f=H_fQ_f, \quad \text{with} \quad H_f=e^{S_f}\quad \text{and} \quad Q_f=e^{-R_f^{(2)}}.  
\end{equation}  
It can be generalized to the $\alpha$-dependent measure of complexity,  
 
$C_f^{(\alpha)}$, which is defined by  
\begin{equation}  
C_f^{(\alpha)}=H_f^{(\alpha)}Q_f, \quad \text{with} \quad H_f^{(\alpha)} = e^{R_f^{(\alpha)}},  
\end{equation}  
that tends to the measure of complexity $C_f$ in the limit $\alpha\rightarrow 1$.  
It satisfies the next properties:\newline  
(i) $C_f^{(\alpha=2)}=1$,\newline  
(ii) $C_f^{(\alpha)}$ is invariant under scaling transformation,   
$f_{\lambda}=\lambda^{D}f(\lambda {\bf r})$, i. e. $C_{f_{\lambda}}^{(\alpha)}=C_f^{(\alpha)}$,\newline  
(iii) taking into account that R\'enyi entropy is a nonincreasing function of 
$\alpha$ \cite{dembo1991}, it is straightforward to see that   
$C_f^{(\alpha)}\geq 1$ for $\alpha<2$,  
$C_f^{(\alpha)}\leq 1$ for $\alpha>2$, and  
 $C_f^{(\alpha)}$ is also a nonincreasing function of $\alpha$.  
  
Let us finally point out that when $\alpha$ goes to $1$,   
the lower bound (that takes the value $1$) for the original LMC complexity  
is recovered \cite{catalan2002}.  
  
\section{Calculations on the hydrogen atom}  
  
The probability density, $\rho({\bf r})$, for a bound state, $\Psi_{nlm}({\bf r})$,  
with quantum numbers $(n,l,m)$ of the non-relativistic H-atom is  
given by $\rho({\bf r})=|\Psi_{nlm}({\bf r})|^2$ in position space (${\bf r}=(r,\Omega)$   
with $r$ the radial distance and $\Omega$ the solid angle), with  
\begin{equation}  
\Psi_{nlm}({\bf r})=R_{nl}(r) Y_{lm}(\Omega).  
\label{wf}  
\end{equation}  
The radial part, $R_{nl}(r)$, is expressed as \cite{galindo1991}  
\begin{equation}  
R_{nl}(r)=\frac{2}{n^2}\left(\frac{(n-l-1)!}{(n+l)!}\right)^{1/2}  
\left(\frac{2r}{n}\right)^l e^{-r/n}L^{2l+1}_{n-l-1}\left(\frac{2r}{n}\right),  
\end{equation}  
with $L_\alpha^\beta(r)$ the associated Laguerre polynomials, and   
$Y_{lm}(\Omega)$ the spherical harmonic of the atomic state.  
Let us recall at this point the range of the quantum numbers:   
$n\geq 1$, $0\leq l \leq n-1$, and $-l\leq m \leq l$.  
Atomic units are used through the text.  
  
Taking the density $\rho({\bf r})$, we have calculated the Fisher-R\'enyi product   
$P_{\rho}^{(\alpha)}$ and the generalized complexity measure $C_{\rho}^{(\alpha)}$ for   
different $(n,l,m)$ wave functions of the $H$-atom.  
  
In Figure 1, the entropy product $P_{\rho}^{(\alpha)}$ was computed numerically for  
$n=15$ and $l=5,10,14$ versus $|m|$ with $\alpha=0.6$ (Fig. 1(a)) and $\alpha=0.8$ (Fig. 1(b)).  
One can see that the minimum of this quantity is given when $l=n-1$.  
This behavior where the energy degeneracy is split  
by $P_{\rho}^{(\alpha)}$  is found for any energy level $n$, and also for any $\alpha$,  
with $0.5<\alpha\leq 1$. 
  
In Figure 2, the value of $C_{\rho}^{(\alpha)}$ is shown for  
$n=15$ and $l=5,10,14$ versus $|m|$ with $\alpha=0.6$ (Fig. 2(a)),  
$\alpha=1.5$ (Fig. 2(b)), and $\alpha=2.5$ (Fig. 2(c)).  
As before, it can be seen that the minimum of $C_{\rho}^{(\alpha)}$ corresponds   
just to the highest $l$ when $0<\alpha<2$ (Figs. 2(a) and 2(b)), but for $\alpha>2$ (Fig. 2(c)),  
we observe the opposite behavior, that is, the highest $l$ presents the maximum value  
of $C_{\rho}^{(\alpha)}$. So, a similar information is provided by both measures   
$C_{\rho}^{(\alpha)}$ when $0<\alpha<2$ and $P_{\rho}^{(\alpha)}$ when $0.5<\alpha\leq 1$.   
  
In Figure 3, the value of $C_{\rho}^{(\alpha)}$ is shown for  
$n=15$ and $l=5,10,14$ (and $0\leq |m|\leq l$) versus $Q_{\rho}$   
with $\alpha=1.5$ (Fig. 3(a)) and $\alpha=2.5$ (Fig. 3(b)). Let us observe in this figure  
the two possibilities of the property (iii) for the quantity $C_{\rho}^{(\alpha)}$.  
In Fig. 3(a), we have $\alpha=1.5$, then $C^{(\alpha)}\geq 1$,  
and in Fig. 3(b), $\alpha=2.5$, then $C^{(\alpha)}\leq 1$.

\section{Summary}

When the Shannon information ingredient of the statistical complexity and the Fisher-Shannon   
entropy is substituted by the R\'enyi entropy of order $\alpha$,  
two $\alpha$-families of statistical measures are obtained.  
Some of their properties have been presented. 
Also, they have been calculated on the hydrogen atom taking advantage   
of the exact knowledge of its wave functions.  
We have put in evidence that, for a fixed level of energy $n$, these quantities,  
the generalized statistical complexity and Fisher-R\'enyi entropy product,  
take their minimum values for the highest allowed orbital angular momentum, $l=n-1$.  
This behavior is displayed when $0<\alpha<2$ for the generalized statistical complexity,  
and when $0.5<\alpha\leq 1$ for the Fisher-R\'enyi entropy product,  
just in the range where uncertainty relations have been shown for this indicator.  
When $\alpha>2$, the opposite behavior is found for the generalized statistical complexity,  
i.e. the maximum of this quantity is taken on the highest angular momentum, and  
for the Fisher-R\'enyi entropy product, the last described behavior is lost when $\alpha>1$.

We conclude by observing that the exchange of the Shannon information by the R\'enyi entropy   
of order $\alpha$ in the statistical complexity and Fisher-Shannon entropy still keeps the   
property of energy degeneracy breaking in the $H$-atom. Further work could be required in order  
to unveil if this feature is displayed in other quantum systems.

\newpage  
\begin{figure}[h]  
\centerline{\includegraphics[width=8cm]{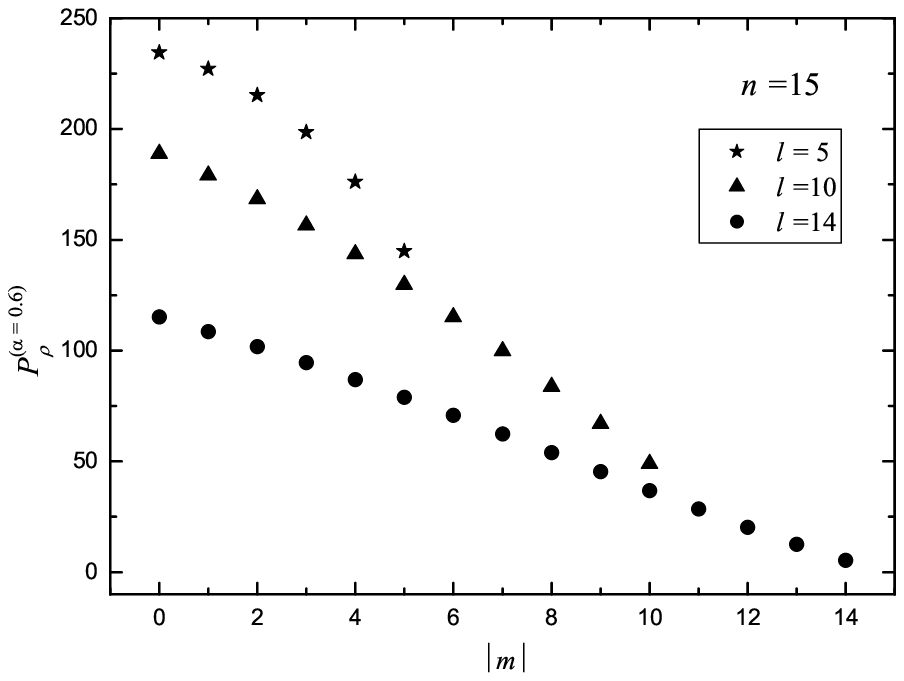}\hskip 5mm\includegraphics[width=8cm]{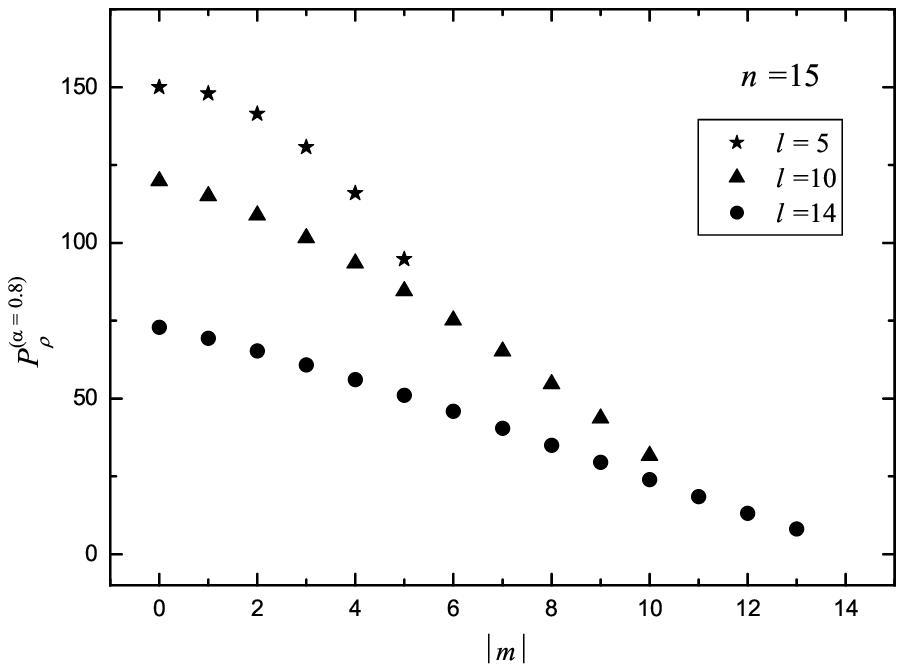}}  
\centerline{(a)\hskip 7cm (b)}   
\caption{Fisher-R\'enyi entropy product in position space, $P_{\rho}^{(\alpha)}$ vs.   
$|m|$ for different $l$ values when $n=15$ in the hydrogen atom. (a) $\alpha=0.6$ and  
(b) $\alpha=0.8$. All values are in atomic units.}  
\label{fig1}  
\end{figure}  
  
\newpage  
\begin{figure}[h]  
\centerline{\includegraphics[width=8cm]{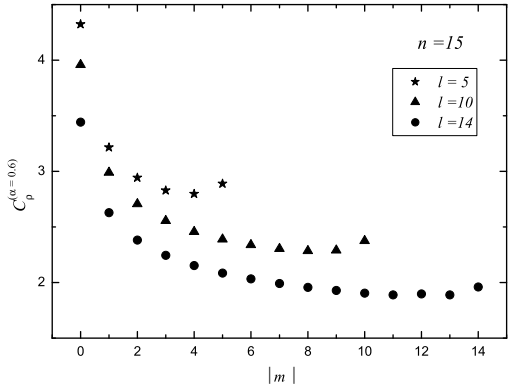}\hskip 5mm\includegraphics[width=8cm]{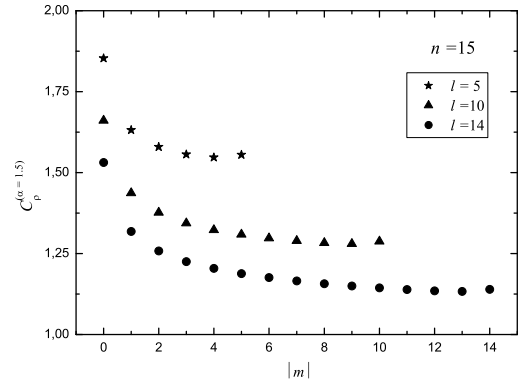}}  
\centerline{(a)\hskip 7cm (b)}   
\centerline{\includegraphics[width=8cm]{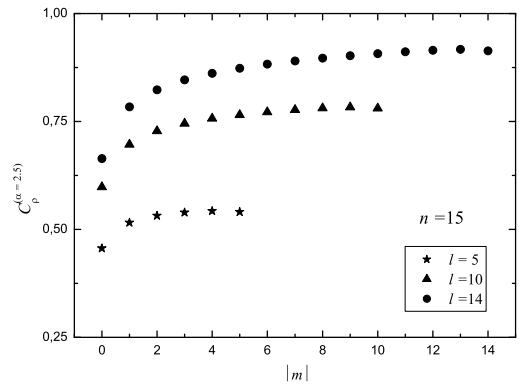}}  
\centerline{(c)}   
\caption{Generalized statistical complexity in position space, $C_{\rho}^{(\alpha)}$ vs.   
$|m|$ for different $l$ values when $n=15$ in the hydrogen atom. (a) $\alpha=0.6$,  
(b) $\alpha=1.5$, and $\alpha=2.5$. All values are in atomic units.}  
\label{fig2}  
\end{figure}  
  
\newpage  
\begin{figure}[h]  
\centerline{\includegraphics[width=8cm]{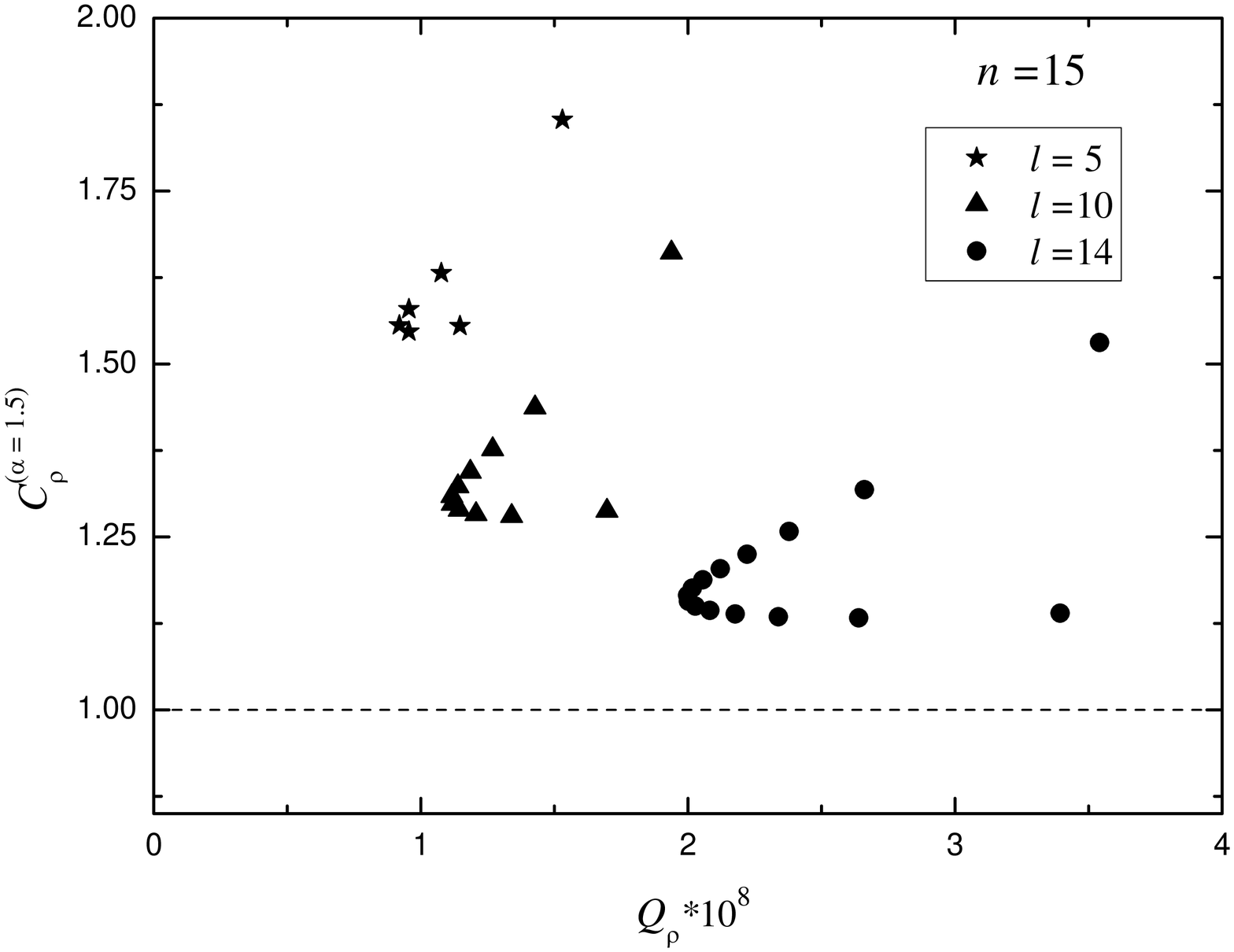}\hskip 5mm\includegraphics[width=8cm]{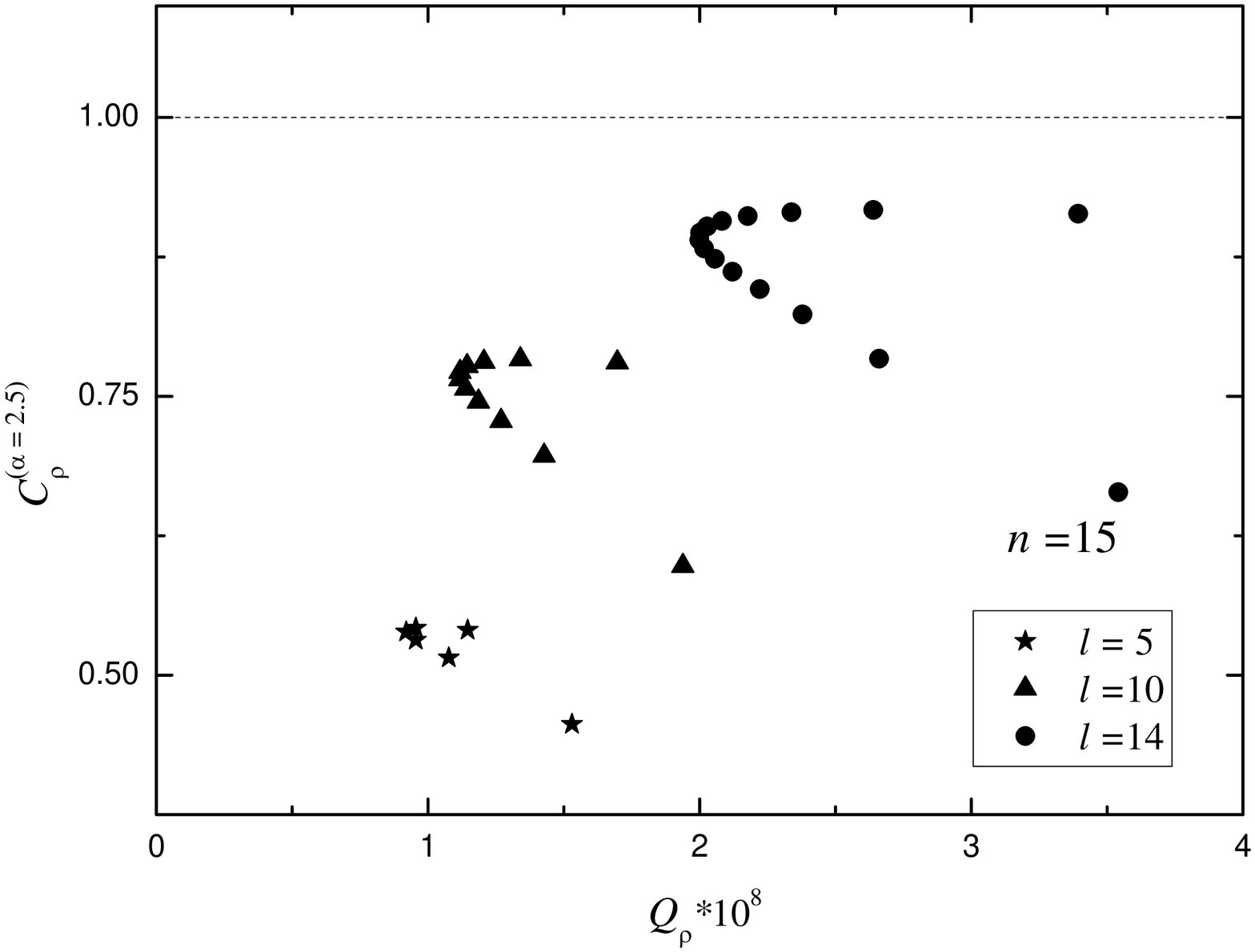}}  
\centerline{(a)\hskip 7cm (b)}   
\caption{Generalized statistical complexity in position space, $C_{\rho}^{(\alpha)}$ vs.   
$Q_{\rho}$ for different $l$ values when $n=15$ and $0\leq |m|\leq l$ in the hydrogen atom.   
(a) $\alpha=1.5$ and (b) $\alpha=2.5$.  
The dashed lines represent the lower bound ($=1$)  
and the upper bound ($=1$) for the $C_{\rho}^{(\alpha)}$, in the cases $0<\alpha<2$ and $\alpha>2$, 
respectively. All values are in atomic units.}  
\label{fig3}  
\end{figure}

\bigskip    
     

\newpage  
\begin{thebibliography}{100}   
   
\bibitem{gadre1985} S.R. Gadre, S.B. Sears, S.J. Chakravorty, and R.D. Bendale,  
Phys. Rev. A {\bf 32} (1985) 2602.   
  
\bibitem{panos2005} K.Ch. Chatzisavvas, Ch.C. Moustakidis, and C.P. Panos,  
J. Chem. Phys. {\bf 123} (2005) 174111.  
   
\bibitem{dehesa1994} R.J. Y\'a\~nez, W. van Assche, and J.S. Dehesa,  
Phys. Rev. A {\bf 50} (1994) 3065.  
   
\bibitem{coffey2003} M.W. Coffey, J. Phys. A: Math. Gen. {\bf 36} (2003) 7441.  
  
\bibitem{romera2005} E. Romera, P. S\'anchez-Moreno, and J.S. Dehesa,  
Chem. Phys. Lett. {\bf 414} (2005) 468.  
 
\bibitem{szabo2008} J. B. Szabo, K. D. Sen, and A. Nagy, Phys. Lett. A {\bf 372} (2008) 2428. 
 
\bibitem{sanudo2008} J. Sa\~nudo and R. L\'opez-Ruiz,  Phys. Lett. A {\bf 372} (2008) 5283.  
 
\bibitem{vignat2003} C. Vignat, J. F. Bercher, Phys. Lett. A {\bf 312} (2003) 27.  
  
\bibitem{romera2004} E. Romera and J. S. Dehesa, J. Chem. Phys. {\bf 120}  (2004) 8906.  
  
\bibitem{angulo2008} J.C. Angulo, J. Antol\'{\i}n, and K.D. Sen,  
Phys. Lett. A {\bf 372} (2008) 670.  
   
\bibitem{lopez1995} R. L\'opez-Ruiz, H. L. Mancini, and X. Calbet, Phys. Lett. A {\bf 209} (1995) 321.  
  
\bibitem{catalan2002} R.G. Catalan, J. Garay, and R. L\'opez-Ruiz, Phys. Rev. E {\bf 66} (2002) 011102.  
  
\bibitem{shannon1948} C.E. Shannon, {\it A mathematical theory of communication},  
Bell. Sys. Tech. J. {\bf 27} (1948) 379; ibid. (1948) 623.  
  
\bibitem{renyi1961} A. R\'enyi, Proceedings of {\it the 4th Berkeley Symposium on Mathematical Statistics   
and Probability}, Volume 1: Contributions to the Theory of Statistics (1961) 547.  
  
\bibitem{tsallis1988} C. Tsallis, J. Stat. Phys. 52, (1988) 479.  
  
\bibitem{rosso2006} M.T. Martin, A. Plastino and O.A. Rosso,   
Physica A {\bf 369} (2006) 439.  
  
\bibitem{romera2008} E. Romera and \'A. Nagy, Phys. Lett. A {\bf 372} (2008) 6823.  
  
\bibitem{dembo1991} A. Dembo, T.M. Cover, and J.A. Thomas, IEEE Trans. Inf. Theor. {\bf 37} (1991) 1501.  
  
\bibitem{fisher1925} R.A. Fisher,  
Proc. Cambridge Phil. Soc. {\bf 22} (1925) 700.  
  
\bibitem{galindo1991} A. Galindo and P. Pascual,  
{\it Quantum Mechanics I}, Springer, Berlin, 1991.  
  
  
\end{thebibliography}
\end{document}